\documentclass[showpacs,twocolumn,prd,nofootinbib]{revtex4-1}
 \usepackage{color,graphicx,epsfig}
 \usepackage{amsmath,amssymb,dsfont,epsfig,graphicx,xcolor}
 \usepackage{ifpdf}
 \usepackage{amsmath}
 \usepackage{bm}
 \usepackage{color}
 \usepackage[english]{babel}
 \usepackage{graphicx}%
 \usepackage{amsfonts}%
 \usepackage{amssymb}
 \usepackage{braket}
 \usepackage{hyperref}
 \usepackage{epstopdf}
 \usepackage{slashed}
 \graphicspath{{Plots/}}
 \usepackage{commath}

\begin{document}

\def\LjubljanaFMF{Faculty of Mathematics and Physics, University of Ljubljana,
 Jadranska 19, 1000 Ljubljana, Slovenia }
\def\LjubljanaIJS{Jo\v zef Stefan Institute, Jamova 39, 1000 Ljubljana, Slovenia}

\title{Coloured Scalar Mediated Nucleon Decays to Invisible Fermion}

\author{Svjetlana Fajfer}
\email[Electronic address:]{svjetlana.fajfer@ijs.si} 
\affiliation{\LjubljanaIJS}
\affiliation{\LjubljanaFMF}
\author{David Susi\v c}
\email[Electronic address:]{david.susic@ijs.si} 
\affiliation{\LjubljanaIJS}
\affiliation{\LjubljanaFMF}

\begin{abstract}
We investigate nucleon decays to light invisible fermion mediated by the coloured scalar $\bar S_1= (\bar 3, 1, -2/3)$ and compare them with the results coming from the mediation of $S_1 = (\bar 3,1,1/3)$. 
In the case of $\bar S_1= (\bar 3, 1, -2/3)$ up-like quarks couple to the invisible fermion, while in the case of $S_1 = (\bar 3,1,1/3)$ the down-like quarks couple to the invisible fermion. For the mass of invisible fermion smaller than the mass $m_p - m_K$, proton (neutron) can decay to $K$ and invisible fermion and the masses of $\bar S_1$ and $S_1$ are in the region $\sim 10^{15}$ GeV. The decays of nucleons to pions and invisible fermion can occur at the tree-level, but in the case of $\bar S_1$ they come from dimension-9 operator and are therefore suppressed by several orders of magnitude compared to the decays into kaons. 
For the invisible fermion mass in the range $(937.8 \, {\rm MeV},\, 938.8 \, {\rm MeV})$, decay of neutron 
$n \to\chi \gamma$ induced by $\bar S_1$ is possible at the loop level, while the proton remains stable. The branching ratio of such decay is $\le 10^{-6}$, which does not explain neutron decay anomaly, but is in agreement with the Borexino experiment bound. We comment on low-energy processes with the nucleon-like mass of $\chi$ in the final state as $\Lambda \to \chi \gamma$ and heavy hadron decays to invisibles. 
\end{abstract}

\maketitle

\newpage

Many constraints on physics beyond the Standard Model (SM) at low-energies are already well established. Although, it seems that possibilities for New Physics (NP) at low energies are known and well studied, there are some chances that light neutral particles may have evaded experiments due to their long lifetime. Recently, the author of Ref. \cite{Heeck:2020nbq} suggested this possibility and investigated a number of scenarios with light fermions carrying lepton or baryon number. In this work we focus on the light fermions carrying baryon number. 
As already summarised by many authors \cite{Heeck:2020nbq,McKeen:2020zni,Cline:2018ami,Grinstein:2018ptl,Jin:2018moh,Keung:2019wpw,Cline:2019fxx,Elor:2018twp,Barducci:2018rlx,Davoudiasl:2014gfa,Berezhiani:2005hv,Berezhiani:2015afa}, 
such interactions between quarks and right-handed fermions are mediated by coloured scalars. Obviously, coloured scalars can couple either to down-like quarks or to up-like quarks depending on their charge $-1/3$ or $2/3$. 
On the experimental side, the KamLAND Collaboration \cite{Abe:2013lua} has already searched for the invisible decays of neutrons, but assumed zero mass of the invisible state.

Leptoquarks mediate SM quark and lepton interactions. In the case where instead of a lepton there is a fermion with quantum numbers of a right handed neutrino, we name the mediator coloured scalar. Following the notation of \cite{Dorsner:2016wpm}, we present in Table \ref{tab:LQ_scenarios} coloured scalars which have interactions with a such state as well as the di-quark interactions.
\begin{table}[h]
\centering
\begin{tabular}{|c|c|c|}
\hline
Cloured Scalar& Invisible fermion & Di-quark \\
\hline

$S_1 = (\bar 3,1,1/3)$ & $ \bar d_L ^{C\, i} \nu_R S_1$ & $\bar{u}_{R}^{C\,i} d_{R}^{j} S^{*}_{1}$\\ 
$\bar S_1= (\bar 3, 1, -2/3)$ & $ \bar u^{C\,j}_R \nu_R \bar S_1$& $ \bar{d}_{R}^{C\,i} d_{R}^{j} \bar{S}_1^{*}$\\
\hline 
\end{tabular}
\caption{\label{tab:LQ_scenarios} The coloured scalars $\bar S_1$ and $S_1$ interactions with invisible fermions and two quarks. Here we use only right handed couplings of $S_1$. Indices $i,j$ refer to quark generations.}
\end{table}
The scalar $S_1$ couples to leptons and therefore plays a role of a leptoquark. Contrary to $S_1(\bar 3, 1 ,1/3) $, $\bar S_1 (\bar 3, 1 ,-2/3) $ is a coloured-scalar (triplet of colour group, singlet of weak, with hypercharge and electric charge equal to $2/3$; here the weak hypercharge $Y$ is defined as $Q=I_3+Y$). Due to its quantum numbers, $S_1$ might have interactions with SM doublets, quarks and leptons, while the coloured scalar $\bar S_1=(\bar 3, 1,-2/3)$ \cite{Dorsner:2016wpm,delAguila:2008ir} only has two type of interactions with right-handed fermions. One with up quarks and with neutral weak right-handed singlets and the second one is an interaction between different generations of the down-quarks \cite{Dorsner:2016wpm}.

In addition to the general study of Ref. \cite{Heeck:2020nbq}, an interesting possibility was discussed in the literature with the main concern being stability of proton, while neutron or hydrogen atom are unstable
\cite{Grinstein:2018ptl,Cline:2018ami,Berezhiani:2018udo,McKeen:2020zni}.
For example, the authors of \cite{Fornal:2018eol} 
 pointed out that there is a discrepancy between the neutron lifetime measured in beam and bottle experiments. This idea initiated new experimental studies which supported discrepancy between the two experimental results \cite{Serebrov:2019kml} on the level of $3.6\sigma$. 
The world average of the bottle experiment according to Particle Data Group (PDG )\cite{Tanabashi:2018oca} is $\tau_n^{bottle} =(880.2 \pm 1.0)$s and $\tau_n^{beam} = (888.0\pm 2.0)$s. 
In Ref. \cite{Fornal:2018eol} this discrepancy was addressed by assuming that neutron can decay to dark matter (DM) and one photon, or two types of DM. 
 In order to avoid proton destabilization, the authors of this proposal suggested that the dark fermion should have mass in the range $m_p - m_e\leq m_\chi \leq m_p+m_e$ (or $937.8 \, {\rm MeV} < m_\chi < 938.8 \, {\rm MeV}$) in the case of neutron decay to DM fermion and $ \gamma$, while the photon energy is in the range $ 0.782 \, {\rm MeV}< E_\gamma < 1.664 \, {\rm MeV}$. The branching ratio for the decay $n\to \chi \gamma$ which explains the neutron lifetime anomaly should be $\sim 10^{-2}$. The selection of this narrow mass window enables the DM to remain stable. Unfortunately, the direct search for the $n\to \chi \gamma$ decay at the level required to explain the neutron lifetime anomaly was unsuccessful \cite{Tang:2018eln}. 
 Another possibility for the DM presence in the nucleon dynamics was offered in \cite{Berezhiani:2018eds} in which the neutron can convert into mirror neutron, its dark partner from parallel mirror sector. 

The approach of \cite{Fornal:2018eol} assumes that a state with quantum numbers of $S_1=(\bar 3, 1,1/3)$ mediates this interaction. 

The fermionic dark matter in this approach is a colour weak singlet, neutral state $(1,1,0)$, which can couple to down-like quarks. 
Recently, the authors of \cite{McKeen:2020zni} questioned a possibility that hydrogen atom is unstable, whereas proton remains stable. They considered a case where the photon is emitted with the energy smaller than the nucleon binding energy inside nucleus. They noticed that the results of the Borexino experiment \cite{Agostini:2015oze} allow the threshold for the detection of electromagnetic energy depositions to be reduced down to $\sim 200$ keV. Using Borexino data \cite{Agostini:2015oze} they found out, that nontrivial constraints arise from the subdominant radiative decay mode. In such a way, we obtain a direct test of scenarios where the neutron mixes with an invisible fermion without the nuclear physics complications. The main message of this study is that Borexino data restrict the branching ratio of the $n \to \chi \gamma$ to be smaller than $10^{-4}$. The existence of heavy neutron stars also gives the strong limits, since $n \to \chi \gamma$ would allow neutron stars to reach masses below the observed ones \cite{Fornal:2020bzz,Elahi:2020urr,Baym:2018ljz,Motta:2018rxp}. 
 The colour scalar or vector mediation in the processes of interactions of the DM with the SM fermions were considered in varieties of the models (see eg. \cite{Blanke:2017tnb,Baek:2017ykw,Jubb:2017rhm,Mohamadnejad:2019wqb,Agrawal:2014aoa}). The new invisible fermion is stable and therefore might be a candidate for the DM. For the kinematic mass of $m_\chi \le m_p$ only DM annihilation channel 
$\chi \chi \to u_i^c u_j^c$ is allowed. This has been widely discussed in literature \cite{Dev:2015uca,Goyal:2016zeh,Allahverdi:2017edd,Blanke:2017tnb,Jubb:2017rhm}. However, the calculated value is smaller than the thermal cross section for DM $3\times 10^{-26}$ cm$^3$ /s$^2$ 
\cite{Dev:2015uca}. 
Such result means that thermal freeze-out leads to over-produced DM and possible scenarios of a non-thermal production mechanism are necessary to explain the observed DM abundance \cite{Dev:2015uca}. Since we consider only phenomenological aspects of the invisible fermion couplings to a coloured scalar and one of the up quarks, we use invisible fermion instead of DM fermion.

In this paper we first write down Lagrangians for $\bar S_1$ and $S_1$ in Sec. \ref{sec1}. Then in Sec. \ref{sec2} we consider decays of nucleons $p,n \to K \chi$ which can occur at tree-level, as well as $p,n \to \pi \chi$. We compare our results with results coming from the mediation of $S_1$. In Sec. \ref{sec3} we discuss decay $n\to \chi \gamma$ due to mediation of $\bar S_1$. The Sec. \ref{sec4} contains a discussion of consequences at low energies. In Sec. \ref{sec5} we summarise our results. 

\section{Interactions of $\bar{S}_1$ and $S_1$}\label{sec1}

The Lagrangian describing $\bar{S}_1=(\overline{\mathbf{3}},\mathbf{1},-2/3)$ interactions is 

\begin{align}
\label{eq:main_b_S_1}
\mathcal{L}_{\bar S_1}\ \supset &+\bar{y}^{\overline{RR}}_{1\,ij}\bar{u}_{R}^{C\,i} \bar{S}_1 \chi^{j}+\bar{z}^{RR}_{1\,ij}\bar{d}_{R}^{C\,i} \bar{S}_1^* d_{R}^{j}+\textrm{h.c.}
\end{align}
This colour scalar does not couple to charged leptons and interacts only with two different down quarks. In principle, in this Lagrangian three species of invisible fermions
$\chi^j \equiv (1,1,0)$ can exist with the quantum number of the right-handed neutrino $\nu_{R}$. In order to simplify the model, we assume that there is only one $\chi \equiv \chi^j$ for $j=1,2,3$ which can couple to the $u$, $c$ and $t$ quarks. In the matrix $\bar{y}^{\overline{RR}}_{1\,ij}$ we then set $j=1$. Strictly speaking, the Lagrangian refers to quarks and invisible fermions in the flavour basis. 
In order to get these fields in the mass basis, one has to perform appropriate rotations (see for details \cite{Dorsner:2012nq}). Since we consider Lagrangians with the right handed fields only, we treat our couplings in (\ref{eq:main_b_S_1}), as they are already in the mass basis. The colour indices are not presented in (\ref{eq:main_b_S_1}).

Note that $\bar{z}^{RR}_{1\,ij}$ is an antisymmetric matrix in any flavour basis, as well as in colour indices (not specified here, but knowing that 
$\bar{d}_{R}^{C\,i} \bar{S}_1^* d_{R}^{j} \rightarrow \epsilon_{\alpha \beta \gamma} \, \bar{d}_{R,\alpha}^{C\,i} d_{R,\beta}^{j}\bar{S}_{1,\gamma}^* $ and $\bar{z}^{RR}_{1\,ij}=-\bar{z}^{RR}_{1\,ji}$).

In some proposals $\chi^j$ is considered to be a Majorana fermion whose mass can be introduced by the mass term $m_\chi \bar \chi^c \chi $. In such scenarios one can simply assign baryon number $B = 2/3$ to $\bar S_1$ and $B=+1$ to $\chi$ \cite{Allahverdi:2017edd}. That means then that the interacting Lagrangian preserves baryon number, while only the Majorana mass term will be source of the baryon number violation.

The full Lagragian for $S_1$ is given in Eq. (9) of \cite{Dorsner:2016wpm}. Here we give only two terms of it, which we later use in our calculations
\begin{align}
\label{eqS_1}
\mathcal{L} _{S_1}\supset &
y^{\overline{RR}}_{1\,ij}\bar{d}_{R}^{C\,i} S_{1} \chi^{j} +z_{1\, ij}^{RR} \bar{u}_{R}^{C\,i} S^*_{1} d_{R}^{j}+\textrm{h.c.}.
\end{align}
Note that the last term can come with the opposite chirality too, which is not the case with $\bar S_1$.

\section{Nucleon decays to pseudoscalar meson and invisible fermion at GUT Scale}\label{sec2}

In \cite{Heeck:2020nbq} the author considers a number of cases with the invisible fermion having nonzero lepton or baryon number. The most general Lagragnian with $\chi$ having baryon number $B=1$ can be written as \cite{Heeck:2020nbq,Davoudiasl:2013pda}
\begin{equation}
\mathcal{L}_\chi = \bar{\chi} (i \slashed{\partial}-m_\chi )\chi+ \left(\frac{u_i d_j d_k \chi_L^c}{\Lambda_{ijk}^2}+ \frac{Q_i Q_j d_k \chi_L^c}{\tilde\Lambda_{ijk}^2} + \textrm{h.c.}\right).
\label{eq:lagrangian_chi}
\end{equation}
Here $\Lambda$ and $\tilde \Lambda$ denote the scales of New Physics (NP). 
We use here notation introduced in \cite{Heeck:2020nbq} and only write the flavour indices, not indicating Lorentz, colour and isospin indices. Assuming baryon number conservation, neutron - anti-neutron oscillations do not occur. 

Integrating out the leptoquark states, $S_1$ or $\bar S_1$, one can straight-forwardly write the effective Lagrangian for the $(u^j,\, d^k,\, d^l,\,\chi)$ interaction (see Figs. 1, 2, 3 and 4)
\begin{equation}
\mathcal{L}_{eff}(\bar S_1) = \frac{\bar{y}^{\overline{RR}}_{1\,j1} \,\bar{z}^{RR}_{1\,kl}}{M_{\bar S_1}^2} \epsilon_{\alpha \beta \gamma} \left( \bar \chi^{C} P_R u_\alpha^j \right) \left( \bar d^{C\,k} _\beta P_R d_\gamma^l \right).
\label{eq:ampl-barS1}
\end{equation}
In the case of $S_1$ one finds 
\begin{equation}
\mathcal{L}_{eff}( S_1) = \frac{{y}^{\overline{RR}}_{11j} \,{z}^{RR}_{1\,kl}}{M_{S_1}^2} \epsilon_{\alpha \beta \gamma} \left(\bar \chi^{C} P_R d_\alpha^j \right) \left( \bar u^{C\,k} _\beta P_R d_\gamma^l \right).
\label{eq:ampl-S1}
\end{equation}
In eqs. (\ref{eq:ampl-barS1}) and (\ref{eq:ampl-S1}) the dimension-6 operators are of the type $u_i d_j d_k \chi_L^c$ in Eq.(\ref{eq:lagrangian_chi}). The last term in Eq. 
(\ref{eq:lagrangian_chi}) can be generated only from the $S_1$ interactions with the left-handed quarks. 

\begin{figure}[!hbp]
\centering
\includegraphics[scale=1]{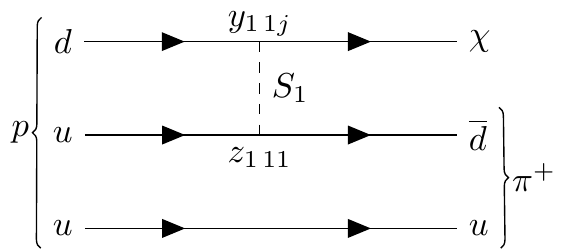}
\caption{ \label{Fig.1} The amplitude for $p \to \chi \pi^+$, induced by $S_1$.}
\end{figure}
\begin{figure}[!hbp]
\centering
\includegraphics[scale=1]{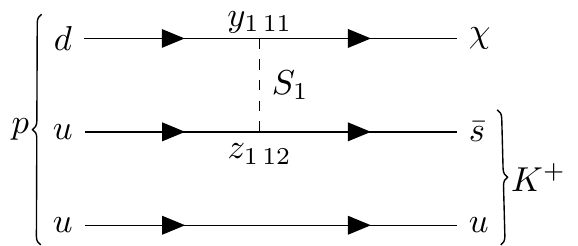}
\caption{ \label{Fig.1} The amplitude for $p \to \chi K^+$, induced by $S_1$. }
\end{figure}
In order to obtain matrix elements of the operator between nucleon and pseudoscalar states one can use notation of Ref. \cite{Aoki:2006ib} 
\begin{align}
< P(p)| \epsilon_{\alpha \beta \gamma} \left( u^T_\alpha C P_\Gamma d_\beta \right) P_{\Gamma^\prime} s _\gamma | N(P,s)>& = \\
\nonumber 
P_{\Gamma}
\left( W_0^{\Gamma \Gamma^\prime} (q^2) -i \slashed{q} W_1^{\Gamma \Gamma^\prime} (q^2) \right) u_N(P,s), 
\label{eq:latt}
\end{align}
with $W_i (q^2)$ being form-factors determined by lattice QCD. One can easily calculate $\bar S_1$ mediated decay amplitudes for $p\to K^+ \chi$ 

\begin{equation} \label{eq:51razpad18}
-i\mathcal{M}=-\frac{i\bar{y}^{\overline{RR}}_{1\,1j}\bar{z}^{RR}_{1\,12}}{m^2_{\bar{S}_1}}W^{RR}_0(k^2_2)\bar{u}_{\chi}(k_2,s_2)P_Ru_p(k_1,s_1)
\end{equation}

\begin{figure}[!hbp]
\begin{center}
\includegraphics[scale=1]{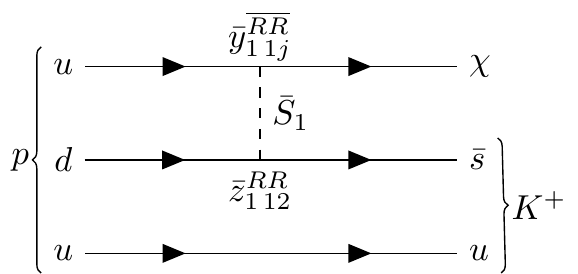}
\caption{Proton decay $p\rightarrow K^+\chi$, induced by $\bar{S}_1$.}
\label{fig:section51diagram1}
\end{center}
\end{figure}
with the decay width 
\begin{equation} \label{eq:51razpad114}
\begin{split}
\Gamma(p\rightarrow K^+\chi)=\frac{1}{32\pi}\left(\frac{\bar{y}^{\overline{RR}}_{1\,11}\bar{z}^{RR}_{1\,12}}{m^2_{\bar{S}_1}}\right)^2|{W^{RR}_0}|^2\\
\times \frac{m^2_p-m^2_{K^+}+m^2_{\chi}}{m^3_p} \lambda^{1/2}(m_{K^+}^2,m_p^2,m_\chi^2) 
\end{split}
\end{equation}
and the decay width for $n \to K^0 \chi$ 
\begin{equation} \label{eq:51razpad21}
\begin{split}
\Gamma(n\rightarrow K^0\chi)=\frac{1}{32\pi}\left(\frac{\bar{y}^{\overline{RR}}_{1\,11}\bar{z}^{RR}_{1\,12}}{m^2_{\bar{S}_1}}\right)^2|{W^{RR}_0}|^2\\
\times\frac{m^2_n-m^2_{K^0}+m^2_{\chi}}{m^3_n} \lambda^{1/2}(m_{K^0}^2,m_n^2,m_\chi^2),
\end{split}
\end{equation}
where $\lambda (a,b,c)= a^2+b^2+c^2- 2( a b+ a c +b c)$. We use results $W^{RR}_{0\,p\rightarrow\pi^+}={0.122}$ GeV$^2$, $W^{RR}_{0\,p\rightarrow K^+}=-W^{RR}_{0\,n\rightarrow K^0}={-0.085}$ GeV$^2$ \cite{Aoki:2006ib,Aoki:2017puj}. For the intermediate $S_1$ one can use above results by making the replacements $\bar{z}^{RR}_{1\,12}\to {z}^{RR}_{1\,12}$, $m_{\bar S_1}\to m_{ S_1}$. 
Experimental results on 
 nucleon decays to invisible fermions only exist for invisible fermion with the negligible mass. The bounds on the lifetimes are $\tau (p\to \pi^+ \nu) > (390\times 10^{30})$ yr~\cite{Abe:2013lua}, $\tau (n\to \pi^0 \nu) > (1100\times 10^{30})$ yr~\cite{Abe:2013lua}, $\tau (p\to e^+ \nu\nu) > (170\times 10^{30})$ yr~\cite{Takhistov:2014pfw}.
As pointed out by the author of \cite{Heeck:2020nbq}, these limits push the scale $m_{\bar{S}_1}$ above $10^{15}\text{ GeV}$. 
For the nucleon decays to pion and invisible fermion induced by $S_1$ one can use (\ref{eq:51razpad21}), replacing $m_K\to m_\pi$, $z_{1\,12} \to z_{1\,11}$. 
 However, the decay amplitude $N\to \pi \chi$ induced by $\bar S_1$ can occur at loop-level or it can appear at tree-level, due to the operator of dimension-9, as explained in detail in \cite{Dorsner:2012nq}.
 \begin{figure}[!hpb]
\begin{center}
\includegraphics[scale=1]{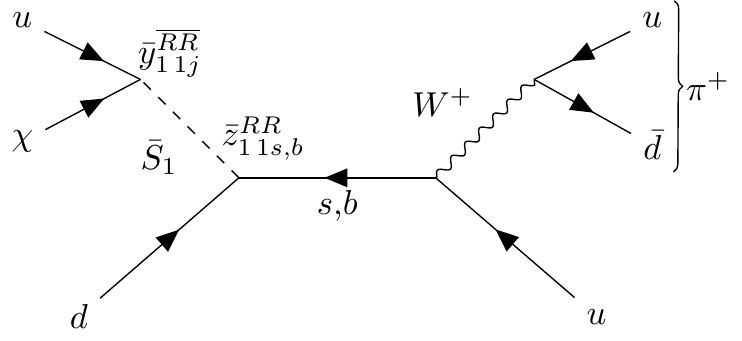}
\caption{Proton decay $p\rightarrow \pi^+\chi$ induced by dimension-9 operator due to $\bar{S}_1$ interaction.}
\label{fig:section51diagram3}
\end{center}
\end{figure}
In Fig. \ref{fig:section51diagram3} the basic decay mechanism caused by the operator of dimension-9 is presented. The effective Lagrangian created by such transition is
\begin{eqnarray}
&&{\cal L}_9 =\frac{8G_F}{\sqrt{2}}\frac{\bar{y}^{\overline{RR}}_{1\,1j}\bar{z}^{RR}_{1\,12}}{m^2_{\bar{S}_1}}\frac{V_{ud}V^{\ast}_{us}}{m_s}\nonumber\\
&&\times \epsilon_{\alpha\beta\delta} (\bar{u}_{\zeta}P_Rd_{\alpha})(\bar{u}^C_{\beta}P_Ld_{\zeta})(\bar{\chi}P_Ru_{\delta}),
\label{Adim9}
\end{eqnarray}
resulting in the amplitude
\begin{eqnarray}
&&\mathcal{M}_{p\rightarrow \pi^+\chi}=i \frac{4G_F}{\sqrt{2}}\frac{\bar{y}^{\overline{RR}}_{1\,1j}\bar{z}^{RR}_{1\,12}}{m^2_{\bar{S}_1}}\frac{V_{ud}V^{\ast}_{us}}{m_s}\nonumber\\
&& \times\frac{f_{\pi^+}m^2_{\pi^+}}{(m_u+m_d)}\alpha_L\bar{u}_{\chi}P_Ru_p
 \label{Amp-9} 
\end{eqnarray}
and the decay width 
\begin{equation} \label{eq:51razpad38}
\begin{split}
\Gamma(p\rightarrow \pi^+\chi)=\frac{1}{4\pi}\left(\frac{\bar{y}^{\overline{RR}}_{1\,11}\bar{z}^{RR}_{1\,12}}{m^2_{\bar{S}_1}}\right)^2\frac{\abs{V_{ud}}^2\abs{V_{us}}^2}{m^2_s}\frac{G^2_Ff^2_{\pi}m^4_{\pi^+}}{(m_u+m_d)^2}\\
\times \alpha^2_L \frac{m^2_p-m^2_{\pi^+}+m^2_{\chi}}{m^3_p} \lambda^{1/2}( m_{\pi^+}^2, m_p^2, m_\chi^2).
\end{split}
\end{equation}
Here the parameter $\alpha_L$ is defined as $P_R u_p \alpha_L = \epsilon^{ijk} <0| \bar u_{Li}^c u_{Lj } d_{Rk} |p> $ (see e.g. \cite{Aoki:2006ib}) with $\alpha_L= 0.0100(12)(214)$ GeV$^3$ obtained by the lattice calculation \cite{Aoki:2017puj}, $f_\pi=0.13$ GeV. 
We do not discuss loop induced $N\to \pi \chi$, due to the additional suppression by the loop factor $1/(16\, \pi^2)$ as explained in \cite{Dorsner:2012nq}. 

It is instructive to determine the suppression factor for the decay widths of $p\to \pi^+ \chi$ and $p\to K^+ \chi$ in the case of $\bar S_1$ with $m_\chi = 0.443$ GeV
\begin{equation}
\frac{\Gamma (p\to \pi^+ \chi) }{\Gamma (p\to K^+ \chi)} \Bigg|_{\bar S_1} \sim 10^{-10},
\label{ratio-bar}
\end{equation}
and in the case of the same processes induced by $S_1$
\begin{equation}
\frac{\Gamma (p\to \pi^+ \chi)}{\Gamma (p\to K^+ \chi)}\Bigg|_{S_1} \sim 10^{-1}. 
\label{ratio}
\end{equation}
In the case of $S_1$ one can derive bound 
\begin{equation}
 \frac{{y}^{\overline{RR}}_{1\, 11} \,{z}^{RR}_{1\,11}}{M_{S_1}^2} \le 2.83 \times 10^{-30} \,{\rm GeV^{-2}}.
 \label{ratio-c}
\end{equation}
In the case of $\bar S_1$, one has the same value for $ {{y}^{\overline{RR}}_{1\, 11} \,{\bar z}^{RR}_{1\,11}}/{M_{\bar S_1}^2}$, 
both determined for $m_\chi= 0.443$ GeV. 
 Obviously, with the improved precision in searches of proton decays, finding $p\to K \chi$ and not seeing $p\to \pi \chi$ would be a possible signature of $\bar S_1$ mediation in nucleon decays. The same processes mediated by $S_1$ 
 does not follow that pattern, differing only by one order of magnitude. 
One might wonder if neutron can decay into pseudoscalar meson and invisible fermion while proton cannot. In the case of kaons in the final state that is not possible due to $m_{K^+} = 0.4937$ GeV being smaller than $m_{K^0}= 0.4976$ GeV. In the pionic case $m_{\pi^+}= 0.13957$ GeV larger than $m_{\pi^0} =0.13497$ GeV. One would think that mass of the invisible fermion should be larger than $m_p - m_{\pi^+}$, which then kinematically forbids the decay $p\to \pi^+ \chi$ and allows $n\to \pi^0 \chi$. However, in both $S_1$ and $\bar S_1$ cases, one can construct the dimension-9 operator which will allow decays $p\to \chi e^+ \nu$ forcing $S_1$ ($\bar S_1$) to have mass of the order of a Grand Unified Theory (GUT) scale. The same mechanism with mass of $m_\chi < m_n - m_\eta$ will imply $n\to \eta \chi$ can occur only at the GUT scale.

\section{ Neutron decays, while  the proton is stable}\label{sec3}

In the case where the mass of invisible fermion is in the range $(937.8 \, {\rm MeV},\, 938.8 \, {\rm MeV})$ proton decay is avoided, but neutron transition to $\chi$ is kinematically allowed. The lower bound on the mass of $\chi$ comes from the request that none of the stable nuclei can decay to dark matter, whereas the upper bound is necessary for the stability of $\chi$ \cite{Fornal:2018eol,McKeen:2020zni,Fornal:2020bzz,Davoudiasl:2014gfa}.
In the case of experimental detection, the simplest way is to register photon of the energy $0.782$ MeV $< E_\gamma < 1.664$ MeV. 
In order to approach the $n \to \chi \gamma$ decay amplitude according to \cite{Fornal:2018eol}, one can assume the mixing of $\chi$ and $n$. Following \cite{Fornal:2018eol}, the effective Lagrangian can be written as
\begin{eqnarray} 
&&{\cal L}_{eff} = \bar n (i \slashed{\partial} - m_n + \frac{g_n e}{8 m_n} \sigma^{\alpha \beta} F_{\alpha \beta}) n \nonumber\\
&&+ \bar \chi (i \partial_\alpha \,\gamma^\alpha - m)\chi +\epsilon(\bar n \chi + \bar \chi n),
\label{eG1}
\end{eqnarray}
where neutron anomalous magnetic moment is $g_n=3.826$ and $\epsilon$ is the mixing parameter with dimension of mass. In the limit $\epsilon\ll m_n -m_\chi$ \cite{Fornal:2018eol}, one easily finds 
\begin{equation}
{\cal L}_{n \to \chi \gamma}^{eff }= \frac{g_n e}{8 m_n} \frac{\epsilon}{m_n- m_\chi} \bar \chi \sigma^{\alpha \beta} F_{\alpha \beta} \, n. 
\label{eG2}
\end{equation} 
In the case considered by \cite{Fornal:2018eol}, the decay $n \to \chi \gamma$ occurs at the tree-level with the mediation of the coloured scalar $(\bar 3, 1, 1/3)$. However, the $\bar S_1$ coloured scalar can mediate such process only at the loop level. Actually, it has to be a box diagram with one $\bar S_1$ and one $W$ (see Fig. 1 ) for the $n\to \chi$ transition. In principle, there is a possibility that in the case of $ u \chi \to s^c (b^c) \bar d$ process, the $s (b)$ quark is transformed to $d$ while the up-like quark and $W$ are mediated in the loop. However, these contributions are suppressed by the mass of $d$ quark and Glashow-Iliopoulos-Miani (GIM) mechanism and can therefore be neglected.

\begin{figure}[!hbp]
\centering
\includegraphics[scale=1]{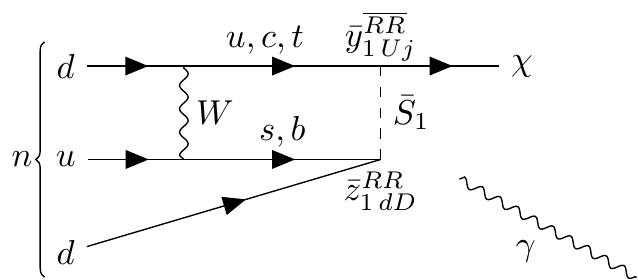}
\caption{ \label{Fig.1} The box diagram contributing to $n \to \chi \gamma$. There are also contributions of the crossed diagram. }
\end{figure}

A contribution of the box diagram to the $n\to \chi$ decay amplitude is presented in Fig. 1. Instead of $\epsilon$, in (\ref{eG1}) we use $\bar \epsilon$ for the mediator $\bar S_1$ 
\begin{equation} \label{eq:51razpad613}
\begin{split}
\bar\varepsilon=\alpha_L\frac{8G_F}{\sqrt{2}}
 \sum_{D=s,b}\sum_{U=u,c,t}\bar y^{\overline{RR}}_{1\,Uj}\bar z^{RR}_{1\,dD} V_{Ud} V^\ast_{uD} \\ \times m_Dm_U I(x_U,x_D,M_{\bar S_1})
\end{split}
\end{equation}
with the integral 
\begin{equation} \label{eq:51razpad614}
\begin{split}
I(x_1,x_2,x_{\bar S_1})=\\
\frac{1}{64 m^2_W\pi^2} \Bigg[\frac{(4-x_1) \, x_1 \, \ln x_1}{(1-x_1) (x_1-x_2)(x_1-x_{\bar S_1})} \\
- \frac{(-4+x_2) \, x_2 \, \ln x_1}{(1-x_1) (x_1-x_2)(x_2-x_{\bar S_1})} \\
+ \frac{(-4+x_{\bar S_1}) \, x_{\bar S_1} \, \ln x_{\bar S_1}}{(1-x_1) (x_1-x_{\bar S_1})(x_2-x_{\bar S_1})} \Bigg].
\end{split}
\end{equation}
In this expression $x_i= m_i^2/m_W^2$.

The dominant contribution from the box diagram comes from the ($c$, $b $) ($t$, $b $) and ($c$, $s$) quarks mediated in the box. 
In the box diagram in Fig. \ref{Fig.1}, the up quarks interact with the coloured scalar $\bar S_1$. The down-like couplings to $\bar S_1$ can be constrained using the oscillations of $K^0 - \bar K^0$, $B^0_{d,s} - \bar B_{d,s}^0$ \cite{Giudice:2011ak,Dorsner:2016wpm}. Note that the couplings in the interacting Lagrangian (\ref{eq:main_b_S_1}) are antisymmetric, which prevents tree-level contributions to these processes. In Appendix A, we present box diagram contributions to the transitions of $K^0 - \bar K^0$, $B^0_{d,s} - \bar B_{d,s}^0$ and determine bounds on the interactions of $\bar S_1$ with the down quarks. 
Here we give bounds on the couplings we use in our calculation: 
$|\bar z^{RR}_{1\, 32 }| \leq 9.21\,({M_{\bar S_1}/\rm{GeV} })^{1/2} 10^{-4} $, $|\bar z^{RR}_{1\, 31 }| \leq 4.18\, ({M_{\bar S_1}/\rm{GeV} })^{1/2} 10^{-3} $ and $|z^{RR}_{1\, 12 }| \leq 0.028\, ({M_{\bar S_1}/\rm{GeV} })^{1/2} $. 

The neutron invisible decay width is given by \cite{Fornal:2018eol} 
\begin{equation}
\Delta \Gamma_{n \to \chi \gamma} = \frac{g_n^2 e^2}{128 \pi} \frac{m_n \bar\varepsilon^2 }{\left(m_n- m_\chi \right)^2} \left(1- \frac{m_\chi^2}{m_n^2}\right)^3.
\label{eG3}
\end{equation}
According to \cite{Fornal:2018eol}, the branching fraction of neutron decay to invisible fermion and photon should be $1\%$ to explain the neutron lifetime anomaly. In their case the parameter is $\epsilon = \beta y^{\overline{RR}}_{1\,11} z^{RR}_{1\, 11}/ m_{S_1}^2$ ($S_1$ corresponds to $\phi$ in \cite{Fornal:2018eol}). The parameter $\beta = 0.0144(3)(21)$ GeV$^3$ \cite{Aoki:2017puj} requires that the branching ratio for $n \to \chi \gamma$ is of the order $1\%$. They obtained that $ y^{\overline{RR}}_{1\,11} z^{RR}_{1\, 11}/ m_{S_1}^2 \sim 8\times 10^{-6}$ TeV$^{-2}$. Note that for $m_{S_1} \sim 1$ TeV the product of $ y^{\overline{RR}}_{1\,11} z^{RR}_{1\, 11} \leq 10^{-6}$. 

In Fig. \ref{Fig.2} we present dependence of the branching ratio $Br(n\to \chi \gamma)$ on the mass of $\chi$ for a given $\bar S_1$ mass. It is interesting that for the mass of $M_{\bar S_1} = 1$ TeV
 the branching ratio is $6.4\times 10^{-7}$, bellow the Borexino limit as discussed in \cite{McKeen:2020zni}. The coloured scalar $\bar S_1$ can have a mass within the TeV regime and is therefore appropriate for the LHC searches. 

\begin{figure}[!hbp]
\centering
\includegraphics[scale=0.8]{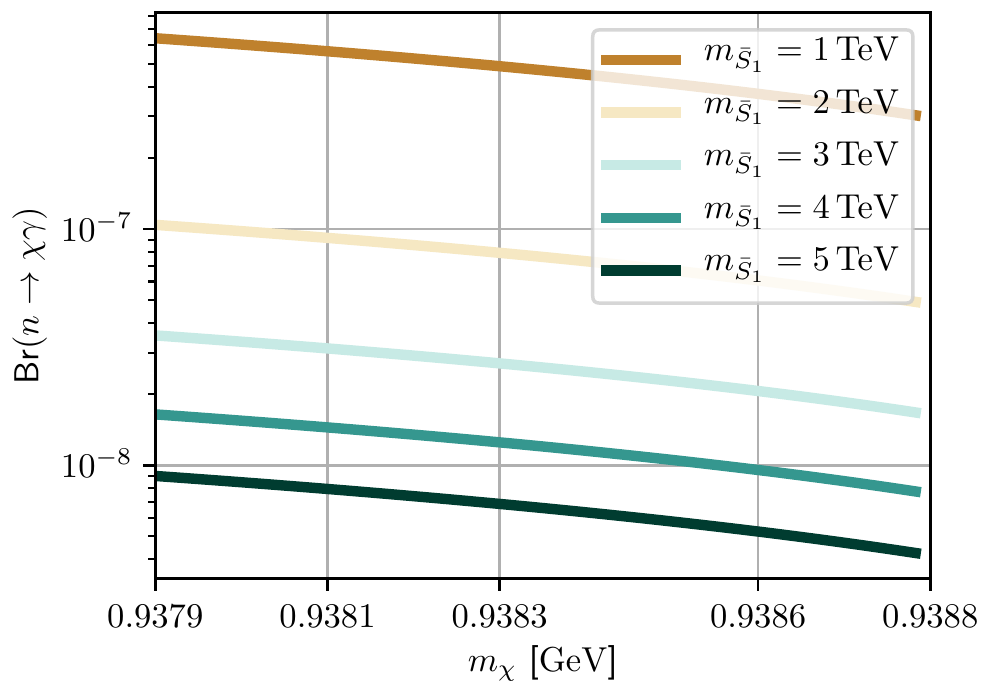}
\caption{ \label{Fig.3} Branching ratio for $n \to \chi\gamma$ as a function of $m_{\chi}$ for the different masses of $\bar S_1$ with $\bar y^{\overline{RR}}_{1\,1j}\simeq {\sqrt 4 \pi}$, $j=2,3$. }
\end{figure}

\begin{figure}[!hbp]
\centering
\includegraphics[scale=0.9]{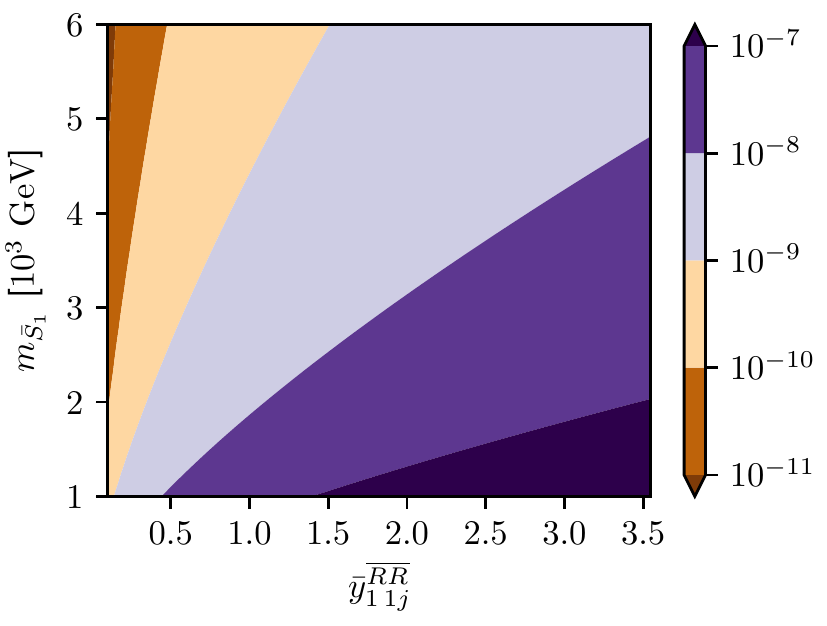}
\caption{ \label{Fig.2} Branching ratio for $n \to \chi\gamma$ as a function of $M_{\bar S_1}$ with $ \bar y^{\overline{RR}}_{1\,12} \simeq \bar y^{\overline{RR}}_{1\,13}$.}
\end{figure}

In Fig.\ref{Fig.3} we present branching ratio dependence or the mass $m_\chi$ GeV and allow the couplings $ \bar y^{\overline{RR}}_{1\,12} \simeq \bar y^{\overline{RR}}_{1\,13} $ to be in perturbative regime. 

\begin{figure}[!hbp]
\centering
\includegraphics[scale=0.8]{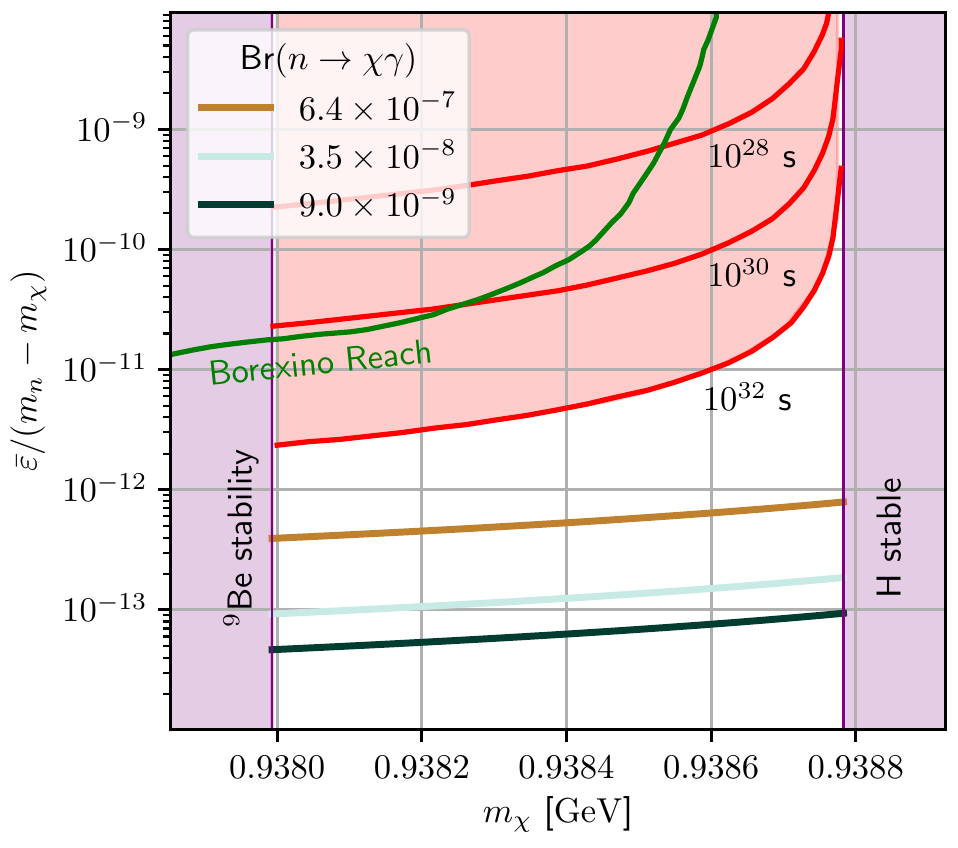}
\caption{ \label{Fig-all} The values $\bar \epsilon/(m_n-m_\chi)$ that yield the neutron decay $n \to \chi\gamma$ as a function of $m_\chi$ for different values of the branching ratio 
$\text{Br}(n\to \chi \gamma)$. The violet regions are excluded by experiment $^9$Be and H stability \cite{McKeen:2020zni}. The red contours indicate atomic hydrogen lifetimes of $10^{28}$s, $10^{30}$s and $10^{32}$s obtained in \cite{McKeen:2020zni} .
}
\end{figure}

The authors of Ref. \cite{McKeen:2020zni} explored the data with expectations of solar neutrinos and backgrounds from radioactivity to derive bounds on the neutron-mixing
parameter $\bar \epsilon/ (m_n-m_\chi)$. They expressed the upper limits on the number of events as lower limits on the H lifetime are $10^{28}$s, $10^{30}$s and $10^{32}$s  (see  Fig. \ref{Fig-all}).The green line is the $90\%$ CL lower limit from their fit procedure to Borexino data. 

The values of parameter $\bar \epsilon/(m_n- m_\chi)$, coming from the calculation of $n-\chi$ oscillations, are allowed by the analysis of \cite{McKeen:2020zni} 
and the mass of $\bar S_1$ can be reached by LHC. In particular, the decay of $\bar S_1$ to two jets and $\bar S_1 \to c (t) \chi$ (monojet) studies were already done by the authors of \cite{Jubb:2017rhm} for larger masses of $\chi$, than the ones we use in this paper.

\section{Possible Low-energy Signatures}\label{sec4}

The processes in which upper quarks couple to an invisible fermion $\chi$ might offer possible experimentally interesting signatures. Here we consider low-energy decays at the tree-level induced by $\bar S_1$ with $\chi$ in the final state. 
These decays have invisible fermions in the final state with mass $m_\chi \simeq 0.938$ GeV, allowed 
by the decay of neutron $n \to \chi \gamma$, leaving the proton stable. We comment on the loop-level decay $b\to s \chi \bar \chi$. 
 The coupling of top quark with $\chi$ and $\bar S_1$ can be nonzero, making a search for $t $ to two jets and invisible particle possible. However, it will be very difficult to distinguish such a signal from the decays of top to two jets at LHC. 

\subsection{$\Lambda\to \chi \gamma$}

\begin{figure}[!hbp]
\centering
\includegraphics[scale=1]{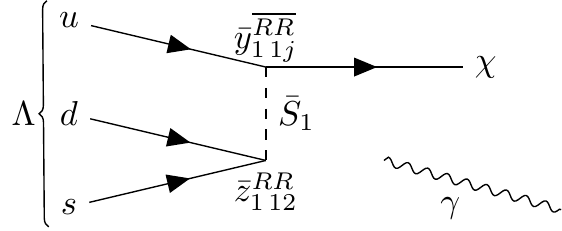}
\caption{ \label{Fig. Lambda} $\Lambda\to \chi \gamma$ .}
\end{figure}

Assuming non-zero coupling of $\chi$ to $u$ quark 
$(\bar y^{RR}_{111} \ne 0)$ one can generate oscillations of the $\Lambda$ baryon to $\chi$ as presented in Fig. \ref{Fig. Lambda}. 
By a simple replacement of $n$ by $\Lambda$ states and $g_n$ by $g_\Lambda$ in equation (\ref{eG3}), one can write 
\begin{eqnarray} 
&&{\cal L}_{eff} (\Lambda) = \bar \Lambda (i \slashed{\partial} - m_\lambda + \frac{g_\lambda e}{8 m_\lambda} \sigma^{\alpha \beta} F_{\alpha \beta}) \lambda \nonumber\\
&&+ \bar \chi (i \partial_\alpha \,\gamma^\alpha - m)\chi) +\epsilon_\Lambda(\bar \lambda \chi + \bar \chi \lambda),
\label{eG2}\end{eqnarray}
leading to the decay width 

\begin{equation}
\Delta \Gamma_{ \Lambda \to \chi \gamma} = \frac{g_ \Lambda^2 e^2}{128 \pi} \frac{m_ \Lambda \bar\varepsilon_ \Lambda^2 }{\left(m_ \Lambda- m_\chi \right)^2} \left(1- \frac{m_\chi^2}{m_ \Lambda^2}\right)^3.
\label{eG4}
\end{equation}
 where $ \bar\varepsilon_ \Lambda= \beta_\Lambda \left( \bar y^{\overline{RR}}_{1\,11}\bar z^{RR}_{1\,12} \right)/M_{\bar S_1}^2$. We use $g_ \Lambda= -1.22$ as given in 
 \cite{Zyla:2020zbs} and assume that the $SU(3)$ flavour symmetry holds. Then, the matrix element $<0| \epsilon_{\rho \sigma \kappa } \left( \bar u_{L\rho}^c d_{R\sigma} s_{R \kappa} \right)| \Lambda> $ is not very different from the matrix element for the neutron, $\beta_\Lambda \simeq \beta = 0. 0144(3)(21)$ GeV$^3$ \cite{Aoki:2006ib}. Current experimental limits on the rates for the baryon number violating processes $\Lambda\to \pi^+ e $ , $\Lambda \to \pi^+ \mu^- $ are smaller than $6\times 10^{-7}$ \cite{McCracken:2015coa,Zyla:2020zbs} and for other searched channel the bounds are even weaker. 
Using Eq. (\ref{eG4}), it is easy to calculate 
\begin{eqnarray}
Br( \Lambda\to \chi \gamma)\bigg|_{M_{\bar S_1}= 5\, {\rm TeV} }&=& |\bar y^{\overline{RR}}_{1\,11}|^2\, 1.75 \times 10^{-6}.
 \label{LBR}
 \end{eqnarray}
 Obviously that such bound would require $\bar y^{\overline{RR}}_{1\,11} \ll 1$. It seems that the coupling of the $u$ quark to the invisible fermion should be very suppressed. From a number of cases studied in the literature (see e.g. \cite{Dorsner:2016wpm,Dorsner:2019vgp} ), the couplings of the first quark generation to leptons and leptoquarks are very suppressed compared to the other two generations. 
 Using the constraint from $D^0- \bar D^0$ oscillations (see Appendix B) we notice, that the product is $\bar y^{\overline{RR}}_{1\,11} \bar y^{\overline{RR} \ast}_{1\,21} < 1.1 \times10^{-5} \, 
 M_{\bar S_1}/{\rm GeV}$. Requirement that $ \bar y^{\overline{RR}}_{1\,11} $ has to be small, leaves a possibility that the coupling $\bar y^{\overline{RR} \ast}_{1\,21} $ can be of the order $1$. This is exactly what is necessary for our analysis of $n \to \chi \gamma$. Obviously, if the $\Lambda \to \chi \gamma$ decay is forbidden, the coupling of $u$ quark should be set to zero. 
 
 \subsection{Heavy hadron decays to invisibles}
 
For the mass $m_\chi \simeq 0.938$ GeV, decays of charmed mesons to invisible fermions are not allowed kinematically. However, baryons containing one $c$ quark and two light quark, e.g. $\Lambda_c^+$ or $\Sigma_c^0$ can decay to invisible fermions. The processes as $\Lambda_c^+ \to K^+ \chi$ and $\Sigma_c^0 \to \chi \gamma$ are allowed. Using Eq. (\ref{eq:51razpad114}), assuming that the matrix element of $<K^+| \epsilon_{\alpha \beta \gamma} \left( c^T_\alpha C P_\Gamma d_\beta \right) P_{\Gamma^\prime} s _\gamma | \Lambda_c^+>$ is not very different from the one in Eq. (6), 
using PDG data for the relevant parameters \cite{Zyla:2020zbs}
we estimate that $Br(\Lambda_c^+ \to K^+ \chi) < 10^{-6}$ (for $\bar y^{\overline{RR} \ast}_{1\,21} \simeq 1$ and $M_{\bar S_1} \sim 2$ GeV). 
The $\Sigma_c$ can decay to $\chi \gamma$. Taking the anomalous magnetic moment of $\Sigma_c^0$ to be $\simeq -2.7$, as calculated in \cite{Riska:1996af}, we obtain, by appropriate replacements in Eq. (\ref{eG3}), that the rate for $\Sigma_c^0 \to \chi \gamma$ is very suppressed, being in the order of $10^{-16}$, making it impossible to be seen. 

Possible decays of heavy hadrons with baryon number violations were discussed in \cite{Hou:2005iu,Elor:2018twp}. The decay $B^+ \to \Lambda_c \chi$ will be allowed within our approach, however very suppressed if the same assumptions as in \cite{Hou:2005iu} are used. 
On the experimental side, there are more searches. For example in BESSIII \cite{Ablikim:2019lil} they search for the processes $D^+ \to \bar \Lambda (\bar \Sigma^0) e^+$ and 
$D^+ \to \Lambda ( \Sigma^0) e^+$, for which the upper limits on the branching fractions are set at the level of $\mathcal O(10^{-6})$. Kinematics forbid 
$D^0 \to \bar \Lambda \chi$ decay for the mass of $\chi$ being close to the nucleon mass. Due to the lack of lattice QCD result on the matrix elements $<\bar \Lambda| \epsilon_{\alpha \beta \gamma} \left( c^\ast_\alpha C P_\Gamma d_\beta ^\dag \right) P_{\Gamma^\prime} s^\dag _\gamma | D^0>$, one can assume they are close in value to the one in Eq. (6). Even if we take $m_\chi =0.443$ GeV, our rough estimate leads to the branching ratio $Br(D^0 \to \Lambda \chi )\le 10^{-19}$, making it too small to be measured.

 \subsubsection{$J/\psi \to \chi \bar \chi$}

The dominant contribution to $\Delta \Gamma(n \to \chi \gamma)$ induced by $\bar S_1$ comes from the coupling of $c$ quark to $\chi$. One would immediately suggest that the $c \bar c$ bound state might decay to two invisible fermions. Only the lower bound $BR(J/\psi \to$ invisibles$) <7 \times 10^{-4}$ is experimentally known.

\begin{figure}[!hbp]
\centering
\includegraphics[scale=1]{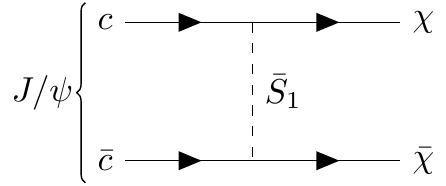}
\caption{ \label{Jpsi2} $J/\psi$ decay to invisibles.}
\end{figure}

The amplitude for decay $J/\psi \to \chi \bar \chi$ at the tree-level, in Fig. \ref{Jpsi2}, can be obtained using the effective Lagrangian approach as in \cite{Dorsner:2016wpm}
\begin{equation}
{\cal L}_{eff} = \sqrt 2 G_F \frac{v^2}{2 M_{\bar S_1}^2} |\bar y^{\overline{RR}}_{1\,12}|^2 (\bar c\gamma_\mu P_R c) (\bar \chi \gamma^\mu P_R \chi).
\label{J-Psi}
\end{equation}

By introducing $< 0 | \bar c \gamma_\mu c| J/\psi (\epsilon, P)> = f_{J\psi} \, m_{J/\psi} \epsilon_\mu$ \cite{Aloni:2017eny}, 
the decay width is given by
\begin{equation}
\Gamma (J/\psi \to \chi \bar \chi) = \frac{f_{J\psi} ^2}{2 \pi m_{J\psi} } (1- 4 x_{\chi}^2)^{1/2} (1- x_{\chi}^2) |{\cal A}|^2
\label{width}
\end{equation}
 with ${\cal A} \equiv\sqrt 2 G_F \frac{v^2}{2 M_{J/\psi}^2} |\bar y^{\overline{RR}}_{1\,21} |^2 $ and $x_{\chi}= m_{\chi}/M_{J/\psi}$.
The experimental bound is very week, allowing huge $\bar y^{\overline{RR}}_{1\,12}$ coupling. For 
$m_\chi=0.938$ GeV and $M_{\bar S_1}$ given in TeV, branching ratio is
\begin{equation}
 Br(J/\psi \to \chi \bar \chi) \leq \frac{|y^{\overline{RR}}_{1\,12}|^4}{M_{\bar S_1}^4} \rm{TeV}^4 \times 10^{-7}.
 \label{BRJpsi}
 \end{equation}
This is three orders of magnitude smaller than the current experimental result in \cite{Zyla:2020zbs}.
 
\subsubsection{$b \to s \chi \bar \chi$}

\begin{figure}[!hbp]

\includegraphics[scale=1]{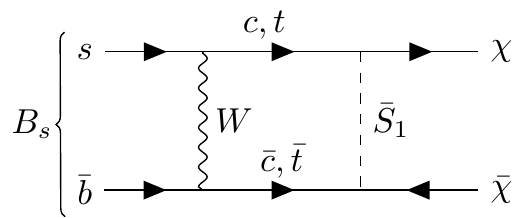}
\caption{ \label{Fig.Bs} $B_s \to \chi \bar \chi$ .}
\end{figure}

The amplitude for $b\to s \chi \bar \chi$ comes from the contributions presented in Fig \ref{Fig.Bs},  and equals to
\begin{eqnarray}
&&{\cal M}( b \to s \chi \bar \chi) = 
\frac{8G_F}{\sqrt{2}}
 \sum_{i,j=c,t} \bar y^{\overline{RR}}_{1\,i1} \bar y^{\overline{RR}\ast}_{1\,j1} V_{ib} V^\ast_{js} \nonumber\\
&& \times m_i m_j (\bar s \gamma^\mu P_L b )\, (\bar  \chi \gamma_\mu P_R \chi )\, I(x_i,x_j,x_{\bar S_1}).
\label{box-bs}
\end{eqnarray}

If we compare the appropriate Wilson coefficient for the $b \to s \chi \bar \chi$ and the numerical value for $M_{\bar S_1}\sim 1$ TeV, we obtain that it is more than two orders of magnitude suppressed compared to the Wilson coefficient for the SM transition $b\to s \nu \bar \nu$ calculated in \cite{Altmannshofer:2009ma}. This makes the invisible fermion search in the exclusive processes $B\to K^{(\ast)} \chi \bar \chi$ very difficult. 
The decays of $B\to K^{(\ast)} \chi \bar \chi$ were considered in Ref. \cite{Li:2020dpc} for the mass of invisible fermions kinematically allowed.
\subsubsection{Possible color scalar signatures at LHC}

A proposal to search for signals of coloured scalars was made in \cite{Abercrombie:2015wmb},
 based on interactions of coloured scalars with up (down) quarks couplings to invisible. 
 The LHC search for coloured scalar with the couplings we consider in this paper, would potentially be performed in the final states containing two light quark jets and mono-jet and missing energy. The authors of Ref. \cite{Pascual-Dias:2020hxo} followed the proposal of Ref. \cite{Giudice:2011ak} and, using the data of \cite{Sirunyan:2018xlo}, derived new bounds for the couplings of colour triplet scalars to two up (down)-like quarks, which were improved by almost two orders of magnitude for light quark jets. However, di-jet couplings are still better constrained by meson oscillations. Hopefully, further LHC searches, such as that of CMS \cite{Sirunyan:2019vgj}, will improve the limits for the model with such particular couplings.

\section{Summary}\label{sec5}

Invisible right-handed fermion can appear in different theoretical frameworks. Here we consider a model in which 
a coloured scalar $\bar S_1 = (\bar 3, 1, -2/3)$ couples either to up-like quarks and invisible right-handed fermion or two down-like quarks of different flavour species.
In the case that both proton and neutron are unstable, decays of $N \to K \chi$ are possible with mass of $\bar S_1$ at GUT scale. The neutron can decay to $n\to \pi^0 \chi$ for the mass of $0.7987$ GeV $< m_\chi< 0.8045$ GeV, while decay $p \to \pi^- \chi$ is forbidden at tree-level by the dimension-9 operator. However, the dimension-9 operator might induce $p\to \chi l^+ \nu_l$ with $l=e, \mu$, forcing the mass of $\bar S_1$ to be at GUT scale. 

In the case when the neutron decays and the proton is stable, the mass of $\chi$ has a very narrow range. The $\bar S_1$ can mediate $n\to \chi \gamma$ at loop-level 
with mass of coloured scalar $\bar S_1$ of the order TeV scale, appropriate for the LHC searches. The contributions of $c$ and $t$ coupling to $\chi$ are largest in this case. The decay rate of $n\to \chi \gamma$ can reach $\sim 10^{-6}$, which is in agreement with the Borexino experiment bound. Further searches of such decays by KamLand and other experiments would help to distinguish between the models of invisible fermions. 
An interesting proposal to search for invisible fermions by their capture by atomic nuclei was done in Ref. \cite{Fornal:2020bzz} suggesting that the large volume neutrino experiments can be used for such searches. This opens up new possibility for searches at DUNE, and at various xenon experiment as explained by the authors \cite{Fornal:2020bzz}. 

Further, we searched for possible signatures of the fermionic invisible particles, coupling to up-quarks via $\bar S_1$ and found that at tree-level one can produce $\Lambda \to \chi 
\gamma$ decay. Obtaining the experimental bound on such decay rate would be very important for the model presented in this paper as well as for obtaining the constraint on the $u$ quark coupling to $\chi$. 
Search for $J/\psi \to \chi \bar \chi$ would shed more light on the possible charm quark coupling to invisible fermions. There are ongoing searches at LHC which will shed more light on the eventual existence  of coloured scalars. 

\section{Acknowledgment} The work of SF was in part financially supported by the Slovenian Research Agency (research core funding No. P1-0035). The work of DS was financially supported by the Slovenian Research Agency (research core funding No. PR-10495).
We are grateful to Darius Faroughy, Damir Be\v cirevi\' c and Nejc Ko\v snik for very insightful discussions.

\section{Appendices}

\subsection{ Di-quark couplings} 

\begin{figure}[!hbp]
\centering
\includegraphics[scale=1]{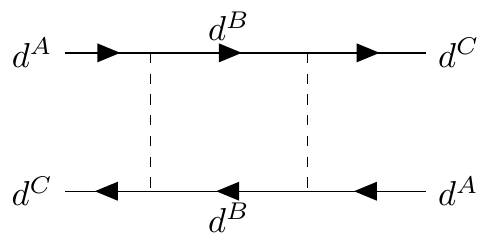}
\includegraphics[scale=1]{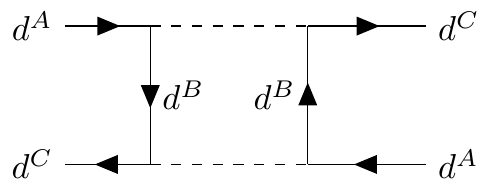}
\caption{ \label{Fig. Osc} The diagrams showing oscillations of mesons consisting of down quark and down anti-quark. The $d^A$ for $A=1,\,2,\, 3$ corresponds $d,\, s,\, b$ quarks. }
\end{figure}

 The contributions from the di-quark couplings in Lagrangian (\ref{eq:main_b_S_1}) appear in the oscillations of $B_s - \bar B_s$, $B_d- \bar B_d$ and $K^0 -\bar K^0$ mesons (see Fig. \ref{Fig. Osc} ).  
 In the case of $B_s - \bar B_s$, there are contributions from the two box diagrams with $d$ quarks within the box. In the case of $B_d- \bar B_d$ ($K^0 - \bar K^0$), internal $s$ ($b$) quarks contribute. 
 The couplings $(\bar z_1)_{ij}$ are antisymmetric ($(\bar z_1)_{ij} = - (\bar z_1)_{ji}$). The contributions of $\bar S_1$ box diagrams in the case of the $B_s - \bar B_s$ oscillation are
 \begin{equation}
 {\cal L}^{NP}_{\Delta B=2} =-\frac{1 }{128 \, \pi^2} \frac{ \left( \bar z^{RR}_{1\, 13 }\right)^2 \left( \bar z^{RR}_{1\, 23 }\right) ^{\ast \, 2} }{M_{\bar S_1}^2} \, \left(\bar s \gamma_\mu P_R b \right) \, \left(\bar s \gamma^\mu P_R b \right).
 \label{deltamBs-S1}
 \end{equation}
 
 This result can be understood in terms of the recent study of new physics in the $B_s - \bar B_s$ oscillation in \cite{DiLuzio:2019jyq}. 
 The authors of \cite{DiLuzio:2019jyq} introduced the following notation of the New Physics (NP) contribution containing the right-handed operators as
 \begin{equation} 
 {\cal L}_{\Delta B=2}^{NP} \supset -\frac{4 G_F}{\sqrt 2} (V_{tb}V_{ts}^*)^2 C_{bs}^{RR} \left(\bar s \gamma_\mu P_R b \right) \, \left( \bar s \gamma^\mu P_R b \right).
 \label{Bs-NP}
 \end{equation}
 Following their notation, one can write the modification of the SM contribution by the NP as in Ref. \cite{DiLuzio:2019jyq}
 \begin{equation}
 \frac{\Delta M_s^{SM+NP}}{\Delta M_s^{SM}}= \left|1 + \frac{\eta^{6/23}}{R_{loop}^{SM}} \, C_{bs}^{RR} \right|
 \label{rationSM-NP}
 \end{equation}
 They found that $R_{loop}^{SM} = (1.31 \pm0.010)\times 10^{-3}$ and $\eta = \alpha_s(\mu_{NP}) /\alpha_s(\mu_b)$. Relying on the Lattice QCD results of the two collaborations FNAL/MILC \cite{Bazavov:2016nty}, HPQCD \cite{Dowdall:2019bea}, the FLAG averaging group \cite{Aoki:2019cca} published following results, which we use in our calculations
 \begin{eqnarray}
\Delta M_s^{FLAG 2019} &= &(20.1^{+1.2}_{-1.6}) \, ps^{-1} = (1.13^{+0.07}_{-0.09}) \, \Delta M_s^{exp} ,\nonumber\\
 \Delta M_d^{FLAG 2019}& = &(0.582^{+0.049}_{-0.056}) \, ps^{-1} = (1.15^{+0.10}_{-0.11}) \, \Delta M_d^{exp} .\nonumber\\
 \label{MBa,e}
 \end{eqnarray}
From these results, one can easily determine bound 
 \begin{equation}
 \frac{ \left( \bar z^{RR}_{1\, 21}\right)^2 \left( \bar z^{RR}_{1\, 31 }\right) ^{\ast \, 2} }{M_{\bar S_1}^2} \leq 1.17 \times 10^{-4}\, {\rm GeV}^{-2},
\label{Bscoef}
\end{equation}
while in the case of $B_d- \bar B_d$, following procedure of \cite{DiLuzio:2019jyq}, by appropriate replacements $s \leftrightarrow d$, the constraint is 
 \begin{equation}
 \frac{\left( \bar z^{RR}_{1\, 21}\right)^2 \left( \bar z^{RR}_{1\, 32 }\right) ^{\ast \, 2} }{M_{\bar S_1}^2} \leq 2.58 \times 10^{-5}\, {\rm GeV}^{-2}.
\label{Bd coef}
\end{equation}
Following work of \cite{Blanke:2011ry,Buras:2013ooa,Agrawal:2014aoa} for the treatment of $K^0 -\bar K^0$, we consider 
\begin{equation}
M_{12}^K = \frac{1}{2m_K} <\bar K^0| {\cal H}_{eff}^{\Delta S=2}|K^0>.
\label{KM12}
\end{equation}
As discussed in \cite{Blanke:2011ry,Buras:2013ooa} the short distance SM value for $M^K_{12}$ 
is found to be 
\begin{equation}
M_{12}^{K,SM} = \frac{G_F^2}{12 \pi^2 }\, f_K^2 \, B_K m_K m_W F_0(x_c,x_t),
\label{KSM}
\end{equation}
with the function $F_0(x_c,x_t)= \lambda_c^2 \eta_{cc} S_0 (x) + \lambda_t^2 \eta_{tt} S_0(y) + 2 \lambda_c \lambda_t \eta_{ct} S_0 (x,y)$. $B_K$ is a bag parameter and $f_K$ is kaon decay constant. They are all introduced in 
\cite{Buras:2013ooa,Blanke:2011ry}.
The effective Lagrangian can be straightforwardly derived by appropriate replacement in Eq. (\ref{Bs-NP}). 

Such Lagrangian gives the following contribution to $M_{K,12}^{\bar S_1} $
\begin{equation}
M_{K,12}^{\bar S_1} = \frac{ \left(\bar z^{RR}_{1\, 31}\right)^2 \left( \bar z^{RR}_{1\, 32 }\right) ^{\ast \, 2 }}{M_{\bar S_1}^2}\, \frac{1}{192 \pi^2} m_K^2 \hat B_K \eta^2.
\label{KNP}
 \end{equation}
The values are $\hat B_K =0.727$, $m_K = 0.4976$ and $\eta_2= 0.58$ as in \cite{Bona:2005eu,Blanke:2017tnb}. This leads to 
 \begin{equation}
 \frac{ \left( \bar z^{RR}_{1\, 31}\right)^2 \left( \bar z^{RR}_{1\, 32 }\right) ^{\ast \, 2 } }{M_{\bar S_1}^2} \leq 3.85 \times 10^{-6}\, {\rm GeV}^{-2}.
\label{Kcoef}
\end{equation}
Using these constraints, one can find 
$|\bar z^{RR}_{1\, 32 }| \leq 9.21 \times 10^{-4} \sqrt{M_{\bar S_1}/\rm{GeV} }$, $|\bar z^{RR}_{1\, 31 }| \leq 4.18 \times 10^{-3} \sqrt{M_{\bar S_1} /\rm{GeV} }$ and $|z^{RR}_{1\, 12 }| \leq 0.028 \sqrt{M_{\bar S_1}/ \rm{GeV} }$ 

\subsection{Constraints from $D^0 - \bar D^0$ }

The effective Hamiltonian describing the $D^0 - \bar D^0$ oscillation is ${\cal H} = C_6 (\bar u_R \gamma C_R) \,(\bar u_R \gamma C_R) $. The effective Wilson coefficient in the case when two $\chi$ and two $\bar S_1$ are exchanged within the box, one can easily calculate
\begin{equation}
C_6(M_{\bar S_1}) = -\frac{ \bar y^{\overline{RR}\, 2}_{1\,11} \bar y^{\overline{RR} \ast \,2}_{1\,21}}{64 \pi^2 M_{\bar S_1}^2}.
\label{c6}
\end{equation}
Usually, the hadronic matrix element $<\bar D^0| (\bar u_R \gamma C_R) \,(\bar u_R \gamma C_R) |D^0> = \frac{2}{3} m_D^2 F_D^2 B$ with the bag parameter 
$B_D(3 GeV) = 0.757(27)(4)$, calculated in the MS scheme, has been computed on the lattice \cite{Carrasco:2015pra}. Due to large nonperturbative contributions, the SM contribution is not well known. Therefore, we can get the robust bound on the product of the couplings by requiring that the mixing frequency, in the absence of CP violation, should be smaller than the world average $x=2|M_{12}|/\Gamma = (0.43^{+0.10}_{-0.11})\%$ as reported by HFLAV \cite{Amhis:2019ckw}. The bound can be obtained as in 
\cite{Fajfer:2015mia} from 
\begin{equation}
|r \,C_6 (M_{\bar S_1})| \frac{ 2 m_D \, f_D^2 B_D}{3 \Gamma_D} < x,
\label{c6bound}
\end{equation}
where $r=0.76$ is a renormalization factor due to running of $C_6$ from scale $M_{\bar S_1} \simeq 1.5$ TeV down to $3$ GeV. One can easily get $|C_6| < 2.2 \times 10^{-13}$ or 
$ \bar y^{\overline{RR}}_{1\,11} \bar y^{\overline{RR} \ast}_{1\,21} < 1.1 \times10^{-5} \, M_{\bar S_1}/{\rm GeV}$.

\bibliographystyle{elsarticle-num}

\bibliography{current1}

\end{document}